\documentstyle[prl,epsfig,aps,amsmath,amssymb,multicol]{revtex}

\begin{document}

\draft

\title{Unexpected Scaling of the Performance of Carbon Nanotube Transistors}

\author{S.~Heinze$^\ddagger$, M.~Radosavljevi\'c$^\ddagger$,
J.~Tersoff$^*$, and Ph.~Avouris$^+$}

\address{IBM Research Division, T.~J.~Watson Research Center,
Yorktown Heights, New York 10598}

\date{\today}

\maketitle

\begin{abstract}
We show that carbon nanotube transistors exhibit scaling that is
qualitatively different than conventional transistors. The
performance depends in an unexpected way on both the thickness and
the dielectric constant of the gate oxide. Experimental
measurements and theoretical calculations provide a consistent
understanding of the scaling, which reflects the very different
device physics of a Schottky barrier transistor with a
quasi-one-dimensional channel contacting a sharp edge. A simple
analytic model gives explicit scaling expressions for key device
parameters such as subthreshold slope, turn-on voltage, and
transconductance.
\end{abstract}


\begin{multicols}{2}

\narrowtext

Recent decades have witnessed remarkable and continuing
improvements in the performance of field-effect transistors
(FETs). These improvements result largely from aggressive scaling
of devices to smaller sizes.  As further improvement of
conventional FETs becomes increasingly difficult, attention has
focused on new devices like carbon nanotube (CN) FETs. CNFETs have
already shown very promising performance, despite the use of
relatively thick gate oxides~\cite{Martel01,Appenzeller02,Shalom02}.

Here we examine the performance improvement of CNFETs upon scaling
of the thickness and dielectric constant of the gate oxide.
In both experimental measurements and numerical calculations,
we find a very different scaling behavior than for conventional
transistors, with important consequences for the design of CNFETs.
Specifically, we find that key measures of device performance
scale approximately as the {\it square root} of the
gate-oxide thickness $t_{ox}$ or its inverse.
These include the turn-on voltage, the transconductance, and the
subthreshold slope.

We show that this surprising behavior can be captured in a simple
analytic model, which gives a universal form for the saturation
current versus gate voltage. Our model incorporates the recent
recognition~\cite{Martel01,Appenzeller02,Dekker01,Heinze02,Nakanishi02}
that, in ambipolar CNFETs such as ours, transistor action is
caused by modulation of the Schottky barriers (SBs) at the
metal-nanotube contacts. The model also highlights the central
role of the contact geometry in determining the scaling, with
different geometries giving different power laws for the scaling.
In contrast, the scaling of conventional FETs with $t_{{\rm ox}}$
is independent of contact geometry in the long-channel limit.

In conventional transistors, there is great interest in oxides
with high dielectric constants, because these increase the gate
capacitance and thus the performance. Improved performance has
also been obtained in this way for
CNFETs~\cite{Appenzeller02,Dai02}. However, for ballistic
SB-CNFETs, the performance is not linked to the capacitance. We
show that in this case, the improvement instead comes from changes
in the electric field patterns due to the {\it inhomogeneous}
dielectric used. For such geometries, the degree of improvement is
tightly coupled to the gate-oxide thickness, with the most
dramatic improvement occurring for thicker gate oxides.

\begin{figure}
\begin{center}
\epsfig{file=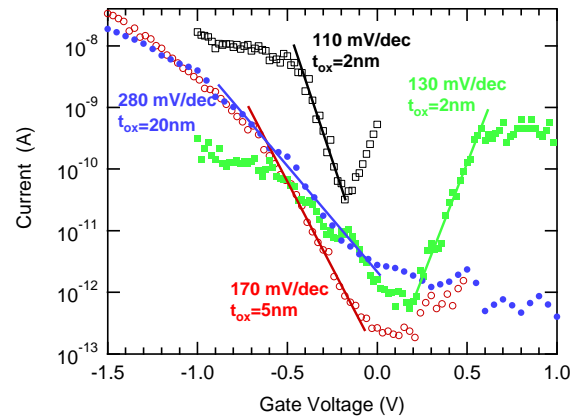,width=8.0cm,angle=0} \caption{
\label{Fig:Exp_I_vs_Vg} Representative transfer characteristics of
CNFETs at different oxide thicknesses $t_{{\rm ox}}$. The turn-on
is characterized by the subthreshold slope (straight lines). The
slope becomes steeper for thinner oxides: from 280~mV/decade for
$t_{{\rm ox}}=20$~nm and $V_{\rm d} =-0.5$~V, to 170~mV/decade for
$t_{{\rm ox}}=5$~nm and $V_{\rm d}=-0.5$~V, and 110 mV/decade for
$t_{{\rm ox}}=2$~nm and $V_{\rm d}=-0.2$~V. Comparable scaling is
seen in n-type devices as well, e.g.\  130~mV/decade for $t_{{\rm
ox}}=2$~nm and $V_{\rm d}=0.2$~V.}
\end{center}
\end{figure}

Our CNFETs use a standard back-gated geometry~\cite{Martel98},
taking advantage of the precise thickness control and high quality
of thermally grown SiO$_2$. Very thin gate oxides (2 and 5~nm) are
grown in small, pre-patterned areas on a degenerately doped
silicon wafer, which serves as the gate. Carbon nanotubes with
diameter of about 1.4nm~\cite{thess_science} are dispersed on the
wafer. Source and drain electrodes, formed using electron-beam
lithography and lift-off, contact those CNs lying on ultra-thin
oxide. In order to suppress the leakage current, the rest of the
source and drain contacts are separated from the back gate by a
thicker ($\sim 120$~nm) field oxide. Details of the fabrication
will be presented elsewhere.

Typical transport characteristics of bottom gate devices with thin
oxides are shown in Fig.~\ref{Fig:Exp_I_vs_Vg}. A quantitative
comparison can be made by measuring the subthreshold slope
\cite{Sze} $S \equiv (d\log_{10}{I}/dV_{\rm g})^{-1}$ where
$V_{\rm g}$ is the gate voltage. The extracted values are given in
Fig.~\ref{Fig:Exp_I_vs_Vg}. We see a steady improvement of the
turn-on for thinner oxides. However, even the thinnest oxides give
$S$ significantly higher than the thermal limit of about $kT \ln
10 \sim$ 60~mV/dec at room temperature, and a key goal here is to
understand the factors leading to further performance improvement.

(Lower $S$ values have been reported in CNFETs using
highly-doped, effectively ohmic contacts~\cite{Dai02} or exotic
designs~\cite{McEuen02}. However, these have their own advantages
and disadvantages; and here we consider only straightforward
scaling of existing designs that lend themselves to integration
with Si-based devices.)

To understand the scaling properties of carbon nanotube (NT)
transistors, we calculate the current using the Landauer-B\"uttiker
formula, and assuming ballistic transport:
\begin{equation}
I=\frac{4e}{h} \int \left[ F(E)-F(E-eV_{\rm d}) \right] T(E) dE ~.
\label{Eq:LB_formula}
\end{equation}
Here $V_{\rm d}$ is the applied drain voltage and $F$ is the Fermi
function. The energy-dependent transmission $T(E)$ through the SB
is evaluated within the WKB approximation, using the idealized
band structure~\cite{White}. This gives
\begin{equation}
\ln{T} = -\frac{4}{3 b V_\pi} \int\limits_{z_{\rm i}}^{z_{\rm f}}
\left(\Delta^2-[E+eV(z)]^2\right)^{1/2} dz,
\label{Eq:Transmission}
\end{equation}
where $b=0.144$~nm is the bond length, $\Delta$ is half the NT
band gap, $V_\pi=2.5$~eV is the tight-binding parameter, and
$V(z)$ is the electrostatic potential along the NT. The
integration is performed between the classical turning points
$z_{\rm i}$ and $z_{\rm f}$. [E.g., for electrons in the
conduction band, $z_{\rm f}$ is determined from $\Delta-eV(z_{\rm
f})=E$; $z_{\rm i}=0$ or, if $E<-\Delta$, is determined from
$-\Delta-eV(z_{\rm i})=E$.] For simplicity, we present numerical
calculations only for the case of mid-gap Schottky barrier, though
we have examined other cases. In this case, the current is
symmetric with respect to the applied gate voltage; and without
loss of generality, we limit the discussion to positive gate
voltages, i.e.\ electron tunneling into the conduction band.

\begin{figure}
\begin{center}
\epsfig{file=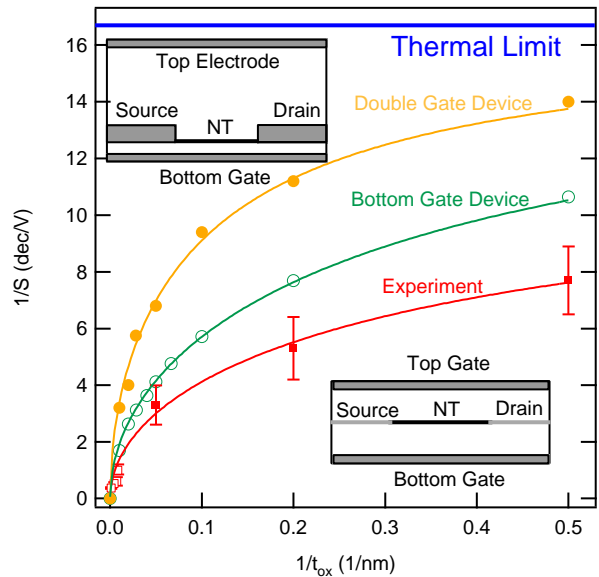,width=8.3cm,angle=0} \caption{
\label{Fig:Slope} Calculated and experimental values of the
inverse subthreshold slope $S$ as a function of the inverse oxide
thickness $t_{{\rm ox}}$. Solid curves are fitted as described in
text. Open red squares are experimental data from previous work,
filled red squares are this work. Data at $t_{{\rm ox}}=2$~nm and
5~nm represent average values for many devices. For $t_{{\rm
ox}}=20$~nm our result agrees with previous reports [3]. The upper
left inset shows the bottom gate device geometry (open green
circles) while the lower right inset shows the double gate device
(filled orange circles).}
\end{center}
\end{figure}

To obtain the electrostatic potential along the NT, we numerically
solve the electrostatic boundary problem given by our device
geometry, which is sketched in the upper inset of
Fig.~\ref{Fig:Slope}. We consider a device similar to the
experiment with a bottom gate at $t_{{\rm ox}}$ from the NT,
source and drain contacts that are 20~nm thick, and a top
electrode which is far from the NT with respect to $t_{{\rm ox}}$.
(The use of a top electrode in this geometry is for computational
convenience; and whether it is grounded or kept at $V_g$ has
negligible impact on the results.) We further neglect charge on
the NT as in~\cite{Heinze02}. This remains a good approximation
throughout the regime studied here~\cite{footnote_charge}. All
numerical calculations here use a NT of 1.4~nm diameter and a band
gap of $0.6$~eV.

In Fig.~\ref{Fig:Slope} the inverse of the subthreshold slope,
$S^{-1}$, has been plotted versus the inverse oxide thickness
$t_{{\rm ox}}^{-1}$.
Experimentally, about 20 to 30
devices were characterized at each oxide thickness.
Only devices with symmetric (or nearly symmetric) p- and
n-conduction were included in the average values presented in
Fig.~\ref{Fig:Slope}.
The $S^{-1}$ values --- even for very thin oxides ---
lie well below the thermal limit.
The theoretical curve for the bottom gate geometry is
in good qualitative agreement with experiment;
the difference presumably arises from the specific
contact geometry, which is not known precisely
on the nm scale in the experiment.

To illustrate the importance of the contact geometry,
we also consider an idealized double gate device
(lower inset of Fig.~\ref{Fig:Slope}).
Its source and drain contacts are as thin as the NT diameter
(1.4 nm), and both top and bottom gate oxides have
thickness $t_{{\rm ox}}$.
The general trend with $t_{{\rm ox}}$ is unchanged,
but for any given $t_{{\rm ox}}$ the subthreshold slope
is greatly improved relative to the bottom gate device
with thicker contact.

\begin{figure}
\begin{center}
\epsfig{file=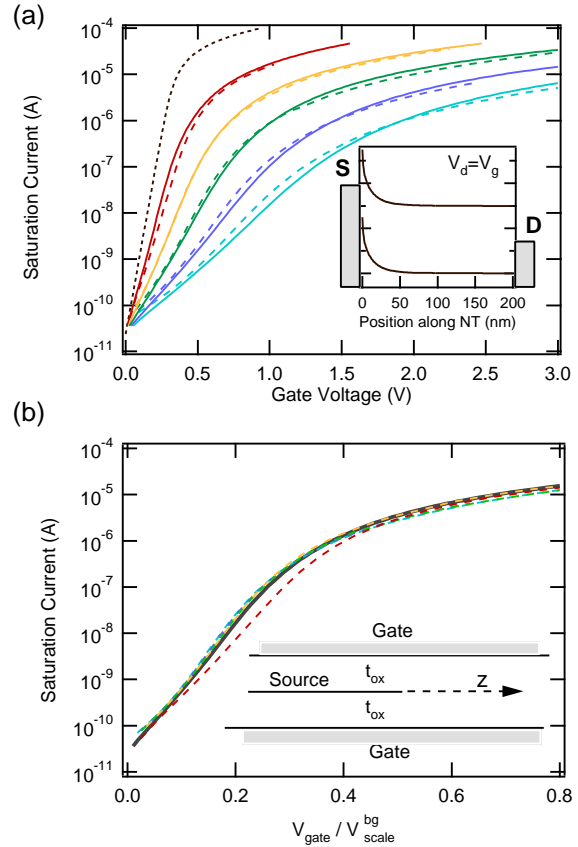,width=8.0cm,angle=0} \caption{
\label{Fig:Scaling} Calculated saturation current versus gate
voltage. (a) shows a set of curves for bottom gate devices with
$t_{{\rm ox}}=35$, 20, 10, 5, and 2~nm, from right to left, and
the thermal limit given by the dotted line. Solid lines are
calculated with Eq.~(\ref{Eq:Analytical_Formula}) and $V_{{\rm
scale}}^{{\rm bg}}=2.2 V_{{\rm scale}}^{{\rm dg}}$ while dashed
lines show the full calculation. (The inset shows the band diagram
for saturating current, i.e.~$V_{\rm d}=V_{\rm g}$.) (b) displays
the same curves versus $V_{\rm g}/V_{{\rm scale}}^{{\rm bg}}$. All
solid lines become one black line,
i.e.~Eq.~(\ref{Eq:Analytical_Formula}). (The inset is a sketch of
the double gate geometry with a semi-infinite sheet contact.)}
\end{center}
\end{figure}

To better understand the behavior, we consider a simple model.
For the double gate device, in the limit of vanishing contact thickness
[inset of Fig.~\ref{Fig:Scaling}(b)],
the electrostatic problem can be solved
analytically~\cite{Morse}. The potential in the vicinity of the
contact varies as $V(z)=2V_{\rm g}\pi^{-1/2} (z/t_{{\rm
ox}})^{1/2}$, where $V_{\rm g}$ is the gate voltage
and $z$ is the distance from the contact along the NT.
Inserting this $V(z)$ into Eq.~(\ref{Eq:Transmission}), and taking
$V_{\rm d}=V_{\rm g}$ [as in inset of Fig.~\ref{Fig:Scaling}(a)],
we calculate the saturation current.
(We consider the limit of a thick gate oxide, where
the position of the conduction band at the drain, $\Delta-eV_{\rm
d}$, can be replaced by $-\infty$).  Then
\begin{equation}
I_{{\rm sat}} = \frac{4e\Delta}{h} \,\, H\left(\frac{V_{\rm
g}}{V_{{\rm scale}}^{{\rm dg}}}\,,\,\frac{\Delta}{kT}\right),
\label{Eq:Analytical_Formula}
\end{equation}
where $H(x,y)$ is
\begin{equation}
H(x,y)=\int\limits_{-\infty}^{\infty}
\frac{\exp{(-h(s)/x^2)}}{1+\exp{(sy)}} \, ds
\label{Eq:Universal_Form}
\end{equation}
and $h(s)$ is
\begin{equation}
h(s)=\int\limits_{\max{(0,-1-s)}}^{1-s} t\, [1-(s+t)^2]^{1/2} \,
dt~. \label{Eq:h}
\end{equation}
The ``scaling voltage'' for this idealized double-gate device is
\begin{equation}
V_{{\rm scale}}^{{\rm dg}}=\left(\frac{2 \pi \Delta^3}{3 b e^2
V_{\pi }}\right)^{1/2} \, t_{{\rm ox}}^{1/2}~. \label{Eq:Vscale}
\end{equation}

Within this idealized geometry (and in the thick-oxide limit etc.),
changing the oxide thickness is equivalent to simply rescaling
the gate voltage by $V_{{\rm scale}}^{{\rm dg}}$.
The turn-on voltage is proportional to $V_{{\rm scale}}^{{\rm dg}}$,
and thus scales as $t_{{\rm ox}}^{1/2}$.
We can evaluate $S$ for $V_{\rm d}$ at saturation, where
$S=(d\log_{10}{I_{{\rm sat}}}/dV_{\rm g})^{-1}$.
Then $S$, like $I_{{\rm sat}}$, is a function of
$V_{\rm g}/V_{{\rm scale}}^{{\rm dg}}$.
Thus $S$ scales as $t_{{\rm ox}}^{1/2}$.
Similarly, we find that the transconductance,
$dI_{{\rm sat}}/dV_{\rm g}$, scales as $t_{{\rm ox}}^{-1/2}$.

For very thin gate oxides, some of these approximations
break down --- in particular, the simple form of $V(z)$ near
the contact, and the infinite limits of integration in
Eq.~(\ref{Eq:Universal_Form}).
But in any case, $S$ must
eventually saturate at the thermal limit of $kT \ln 10 \sim$ 60~mV/dec.
This suggests
the interpolation formula $S=(\alpha t_{{\rm ox}} +
(kT \ln 10)^2)^{1/2}$, with $\alpha$ a fitting parameter.
Fig.~\ref{Fig:Slope} shows that this fitting
works well for the experiment and the calculation for both device
geometries. For thick oxides, $S$ scales as $t_{{\rm
ox}}^{1/2}$, while for very thin oxides it gradually
approaches the thermal limit.

For a more complete description of the device behavior,
we examine the saturation current vs.\  gate voltage.
Figure \ref{Fig:Scaling}(a) shows the results of the
full numerical calculation for the bottom-gate geometry.
These are well described by the analytic model (\ref{Eq:Analytical_Formula}),
over a large range of $t_{{\rm ox}}$,
if we replace the analytic $V_{{\rm scale}}^{{\rm dg}}$ with
$V_{{\rm scale}}^{{\rm bg}}=2.2 V_{{\rm scale}}^{{\rm dg}}$,
where the single empirical factor of 2.2 is sufficient to
account for the difference in geometry.
Figure \ref{Fig:Scaling}(b) shows that all curves are
nearly identical to the analytic form (\ref{Eq:Analytical_Formula}),
with the same $t_{{\rm ox}}^{1/2}$ scaling as in the
idealized model (aside from the one empirical factor of 2.2).
As discussed above, the analytic model becomes inaccurate for
very thin oxides, and this
is observed in Fig.~\ref{Fig:Scaling}(b) for $t_{{\rm ox}}=2$~nm.

In contrast to conventional FETs, the scaling of the performance
here with
$t_{{\rm ox}}$ depends on the specific contact geometry.  If
the contact were infinitely thick, then the electrostatic potential
near the right-angle corner would
depend on distance as $z^{2/3}$~\cite{Morse}.  For such
a geometry, the turn-on voltage and $S$ would scale
as $t_{\rm ox}^{2/3}$, while the transconductance would
scale as $t_{\rm ox}^{-2/3}$.
Thus there is not really a universal power law;
rather, the behavior depends on the contact geometry.
Nevertheless, the two geometries considered here are rather well
described by a single simple power-law behavior.

In addition to reducing the oxide thickness, the performance can
be improved by using an oxide (or other gate dielectric) having
higher dielectric constant \cite{Appenzeller02,Dai02,McEuen02}. In
conventional transistors, this improves performance by increasing
the gate capacitance, and hence the charge in the channel.
However, in ballistic SB-CNFETs, the charge in the channel is
unimportant for the turn-on regime, so the improvement observed in
such devices~\cite{Appenzeller02} must have a different origin.

In fact, within the approximations made here, simply increasing
the dielectric constant everywhere would have {\it no effect} on
the performance. However, most actual devices studied use a thin
gate oxide below the NT, with air above it.
\begin{figure}
\begin{center}
\epsfig{file=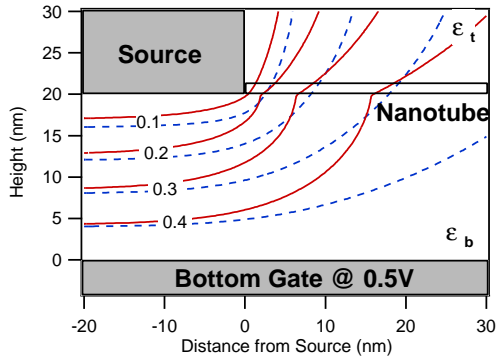,width=7.cm,angle=0} \caption{
\label{Fig:Interface} Contour lines of the electrostatic potential
at the source contact for an interface between a high dielectric
and vacuum (solid red lines, $\epsilon_{\rm b}/\epsilon_{\rm
t}=11$) and a homogeneous dielectric (dotted blue lines). The
value of adjacent contour lines differs by $0.1$~V. ($V_{\rm
d}=V_{\rm g}=0.5$~V.)}
\end{center}
\end{figure}
Fig.~\ref{Fig:Interface} shows how the electrostatic potential
changes due to the interface between dielectrics. The potential
contour lines at the source contact are much closer in the case of
an interface with $\epsilon_{\rm b}>\epsilon_{\rm t}$, resulting
in a thinner SB. Thus the turn-on becomes much sharper.

In this geometry the improvement is large only for relatively
thick oxides. This can be understood from the interface boundary
condition $\epsilon_{\rm b} \partial \phi_{\rm b} / \partial y =
\epsilon_{\rm t}
\partial \phi_{\rm t} / \partial y$, where $\phi_{{\rm t,b}}$ is the
electrostatic potential on top of the interface or below it, and
$\epsilon_{\rm t,b}$ are the dielectric constants of the top and
bottom oxide. In the limit of large dielectric constant for the
bottom oxide, $\epsilon_{\rm t}/\epsilon_{\rm b} \rightarrow 0$,
giving equipotential lines perpendicular to the interface.
However, a distance on the order of the oxide thickness is needed
for a drop of $V_{\rm g}$ of the NT potential. Therefore, while
the turn-on becomes sharper, it does not approach the limit
$T(E)\approx 1$.

Our numerical calculations indicate that with high ratios
$\epsilon_{\rm b}/\epsilon_{\rm t}$, large improvements can be
achieved for thick oxides. However, the changes become rather
small for very thin oxides. For example, for $t_{{\rm ox}}=20$~nm,
replacing the homogeneous oxide ($\epsilon_{\rm b}/\epsilon_{\rm
t}=1$) by HfO$_2$ below and air above the NT ($\epsilon_{\rm
b}/\epsilon_{\rm t}=11$) changes the slope $S$ from 240~mV/dec to
120~mV/dec. This improvement agrees well with experimental
data~\cite{Appenzeller02}. For $t_{{\rm ox}}=2$~nm, though, $S$
changes only from 95~mV/dec to 75~mV/dec.

We have focused here on the case of a mid-gap Schottky barrier.
However, we have also examined the behavior for other values of
the Schottky barrier. As long as the current is limited by
tunneling through a SB, as for barrier heights of $0.2$~eV or
more, the scaling behavior derived here continues to hold.

We thank Bruce Ek for expert technical assistance and J\"org
Appenzeller for helpful discussion. S.~H.~thanks the Deutsche
Forschungsgemeinschaft for financial support under the Grant
number HE3292/2-1.

\end{multicols}


\begin{references}
\bibitem{footnote_charge} The approximation holds whenever
either the source or drain potential barrier (with respect to the
channel, at given gate and drain voltages) is on the order or
smaller than $kT$.
\end{references}
\end{document}